\documentclass[aps,prd,a4paper,preprintnumbers,superscriptaddress,twocolumn,floatfix,final]{revtex4}

\usepackage{amsmath}
\usepackage{graphicx}
\usepackage{amssymb}

\begin{document}
\title{Stabilization of Linear Higher Derivative Gravity with Constraints}

\author{Tai-jun Chen}
\email{T.Chen@damtp.cam.ac.uk}
\affiliation{DAMTP, University of Cambridge, Wilberforce Road, CB3 0WA, Cambridge}
\author{Eugene A. Lim}
\email{eugene.a.lim@gmail.com }
\affiliation{Department of Physics, King's College London, Strand, London, WC2R 2LS}

\date{\today}

\begin{abstract}
We show that the instabilities of higher derivative gravity models with quadratic curvature invariant $\alpha R^2+\beta R_{\mu\nu}R^{\mu\nu}$ can be removed by judicious addition of constraints at the quadratic level of metric fluctuations around Minkowski/de Sitter background. With a suitable parameter choice, we find that the instabilities of helicity-0,  1,  2 modes can be removed while reducing the dimensionality of the  original phase space. To retain the renormalization properties of higher derivative gravity, Lorentz symmetry in the constrained theory is explicitly broken.
\end{abstract}
\maketitle

\section{introduction}
It is well known that \emph{non-degenerate} higher derivative theories suffer from Ostrogradski's instability \cite {ostro, Simon:1990ic, deUrries:1998bi, Woodard, Chen:2012au}. For example,  consider an action with a quadratic 2nd order time derivative term
\begin{equation}
S = \int dt~\left(\frac{1}{2}\ddot{q}^2 - V(q)\right). \label{eqn:S1}
\end{equation}
The equation of motion is 4th order, and hence its phase space is 4 dimensional. We can define the  two canonical coordinates and their conjugate momenta to be $(Q_1, P_1)$, $(Q_2, P_2)$, and the Hamiltonian is hence
\begin{equation}
H = P_1Q_2+\frac{P_2^2}{2}+V(Q_1),
\end{equation}
while $P_1$ only appears linearly in the Hamiltonian as $P_1 Q_{2}$. The linearity of the $P_1$ in this term renders the Hamiltonian unbounded from below and the theory is thus unstable. Because of this undesirable property, non-degenerate higher derivative theories are often viewed as taboo and avoided in the literature.

There are several classes of higher derivative theories in the market, which evade this instability. A higher derivative theory may be \emph{degenerate}, which means that the theory is constrained. For example, in  $f(R)$ gravity \cite{Starobinsky:1980te, DeFelice:2010aj, Sotiriou:2008rp}, the naive unstable degree of freedom is rendered harmless by a gauge constraint. Furthermore, some theories are secretly 2nd order despite the appearance of higher derivative terms in the action due to a clever cancellation of the higher derivative terms in the equation of motion -- as seen in the Galileon theory \cite{ Nicolis:2008in,Deffayet:2009wt,Nicolis:2009qm}.

On the other hand, generic non-degenerate higher derivative theories are inevitably unstable. Theories with curvature invariants such as $R_{\mu\nu}R^{\mu\nu}$, $R_{\mu\nu\sigma\rho}R^{\mu\nu\sigma\rho} $\cite{Stelle:1976gc, Whitt:1984pd, Hindawi:1995an, Hindawi:1995cu,Carroll:2004de,Nunez:2004ts,Chiba:2005nz,Calcagni:2005im,Navarro:2005da,DeFelice:2006pg}, or the Weyl invariant $C_{\mu\nu\sigma\rho}C^{\mu\nu\sigma\rho}$ \cite{Maldacena:2011mk,Lu:2012xu} \footnote{In 4D, the Weyl invariant $C_{\mu\nu\sigma\rho}C^{\mu\nu\sigma\rho}$ can be written as $\frac{1}{2}(R_{\mu\nu}R^{\mu\nu}-\frac{1}{3}R^2)$because the Gauss-Bonnet term $\sqrt{-g}(R_{\mu\nu\sigma\rho}R^{\mu\nu\sigma\rho}-4R_{\mu\nu}R^{\mu\nu}+R^2)$ a total divergence and so does not contribute to the classical equations of motion.} suffer from the sickness of Ostrogradski's instability.

One way to deal with the instability is to impose boundary conditions in such a way that the unstable modes vanish. For example, in \cite{Maldacena:2011mk,Park:2012ds} the modes with the wrong sign of the kinetic terms are ``turned off'' by imposing suitable boundary conditions.  However, this is only valid at the quadratic level. In the presence of higher order interaction terms beyond the quadratic power of the field, the \emph{vacuum} states will rapidly decay (even classically) into states with positive energy modes and negative energy modes by the entropic argument \cite{Kallosh:2007ad, Eliezer:1989cr, Woodard, Chen:2012au}. The ``removed" instability is thus revived.

We will consider the following action first investigated by Stelle \cite{Stelle:1976gc}\footnote{Here we have turned on the bare cosmological constant since the theory admits constant curvature background solution with $R_{\mu\nu}=\Lambda g_{\mu\nu}$.}
\begin{equation}\label{fullaction}
S=\frac{M_{P}^2}{2}\int d^4x~\sqrt{-g}(R-2\Lambda+\alpha R^2+\beta R_{\mu\nu}R^{\mu\nu}).
\end{equation} 
This action with mass dimension $-2$ parameters $\alpha$ and $\beta$ in general contains eight degrees of freedom \cite{Stelle:1977ry}, two of them corresponding to the massless graviton in general relativity, five corresponding to the massive graviton, and the last one is a massive scalar.  Among them, the helicity-2 sector is a non-degenerate higher derivative theory and thus suffers from Ostrogradski's instability.

Nevertheless, this action is interesting as it is power-counting renormalizable \cite{Stelle:1976gc} -- the presence of higher derivative terms in the action means that there exist higher \emph{spatial} derivatives in the propagator of the graviton modes. These spatial derivatives suppress the UV divergences in the loops, rendering the theory naively renormalizable. The price we pay for this is the presence of the higher \emph{time} derivative terms which leads to Ostrogradski's instability. 

One way to take advantage of this insight is to impose different scaling dimensions to the time and space coordinates -- a stratagem utilized by Ho\u{r}ava \cite{Horava:2008ih, Horava:2009uw, Horava:2009if,Mukohyama:2010xz,Sotiriou:2010wn}. The low energy limit of this theory is then a generic 1st order time derivative graviton action with higher order Lorentz violating spatial derivative terms, which is both stable and power-counting renormalizable. 

In this paper, we pursue a different tack. We ask whether we can selectively remove the linear instability by imposing constraints on the theory. This idea is motivated by our recent proof \cite{Chen:2012au} that the linearly unstable phase space can be excised from the theory by a judicious choice of additional constraints (i.e. the final dimensionality of the phase space will be smaller). We will show that, at least in the linear theory, we can add by hand to the theory additional constraint terms  which will render the theory stable, while simultaneously preserving the improved renormalizable features of it. Roughly speaking, we add a constraint where the higher time-like derivative terms in the equation of motion is constrained to some lower time-like derivative or higher order spatial derivative term, i.e.
\begin{equation}
g^{(4)}\sim \partial^2 \ddot{g}, \partial^4 g, \cdots.
\end{equation}
We will show that the final form of this constrained theory is, at least linearly, that of a second order equation of motion of higher order spatial derivatives very similar in spirit to the Ho\u{r}ava model. Of course, such addition of constraints changes the general theory -- however, as we have simply worked in linear theory, we do not know what is the non-linear completion of the theory. We will leave this for future work. 

Our strategy is as follows. In section \ref{sect:linear} we show how to perturb the action up to second order in metric perturbation in general background, which will be used in Minkowski/de Sitter backgrounds. In section \ref{sect:MK} we obtain the action quadratic in the metric fluctuation by parameterizing the metric fluctuation in Minkowski background. Since up to quadratic order, the action can be separated into helicity-0, 1, 2 sectors, we  demonstrate how the instabilities appear in each sector. In section \ref{sect::exorcise}, we show that, how the helicity-0, 1, 2 instabilities can be rendered stable by introducing suitable constraints. We study the behavior and how to remove the instabilities in de Sitter background in section \ref{sect::dS}, \ref{sect::exorcisedS}. We conclude in section \ref{sect:conclusion}.

\section{Higher Derivative Gravity: Quadratic Action}\label{sect:linear}
In order to study how do the instabilities appear in action (\ref{fullaction}) at the quadratic order in the metric fluctuation, we will need to expand every curvature invariant up to second order in the metric perturbation $h_{\mu\nu}$, which is defined by
\begin{equation}\label{pertmetric}
g_{\mu\nu}=\bar{g}_{\mu\nu}+h_{\mu\nu},
\end{equation}
where $\bar{g}_{\mu\nu}$ at this stage can be general background metric and $h_{\mu\nu}\ll \bar{g}_{\mu\nu}$ \cite{Gullu:2010em}. The inverse metric up to second order in $h$ can be written as
\begin{equation}\label{pertinvmetric}
g^{\mu\nu}=\bar{g}^{\mu\nu}-h^{\mu\nu}+h^{\mu\rho}h^{\nu}_{\rho}+O\left(h^3\right).
\end{equation}
Assuming a constant curvature background of either Minkowski ($\Lambda=0$), de Sitter ($\Lambda>0$), or Anti-de Sitter ($\Lambda<0$), we compute the second order action 
\begin{align}\label{action}
S=-\frac{M_P^2}{4}\int d^4x~&\sqrt{-\bar{g}}h^{\mu\nu}\left[(1+8\alpha \Lambda +\frac{4}{3}\beta \Lambda)\mathcal{G}_{\mu\nu}^L\right. \nonumber \\
&\left.+(\beta+2\alpha)(\bar{g}_{\mu\nu}\Box-\bar{\nabla}_\mu\bar{\nabla}_\nu+\Lambda\bar{g}_{\mu\nu})R_L\right.\nonumber \\ 
&\left.+\beta\left(\Box\mathcal{G}^L_{\mu\nu}-\frac{2\Lambda}{3}\bar{g}_{\mu\nu}R_L\right)\right],
\end{align}
where $\Box$ is d'Alembert operator and the linearized Ricci tensor, Ricci scalar, and Einstein tensor are defined by \footnote{see, for example, \cite{Deser:2002jk}}
\begin{align}
R^L_{\mu\nu}&=\frac{1}{2}(\bar{\nabla}_{\rho}\bar{\nabla}_{\mu}h^{\rho}_{\nu}+\bar{\nabla}_{\rho}\bar{\nabla}_{\nu}h^{\rho}_{\mu}-\Box h_{\mu\nu}-\bar{\nabla}_{\mu}\bar{\nabla}_{\nu}h),\nonumber \\
R_L&=\bar{g}^{\mu\nu}R^L_{\mu\nu}-\bar{R}^{\mu\nu}h_{\mu\nu},\nonumber \\
\mathcal{G}_{\mu\nu}^L&=R^L_{\mu\nu}-\frac{1}{2}\bar{g}_{\mu\nu}R_L-\Lambda h_{\mu\nu}.
\end{align}
Note that the indices are raised and lowered by background metric $\bar{g}_{\mu\nu}$.

\section{Quadratic action around Minkowski background}\label{sect:MK}
In this section we want to study how the instabilities appear in the action at the quadratic level of perturbation around Minkowski background $\Lambda=0$. We  parameterize the metric fluctuation by 
\begin{equation}\label{metric}
ds^2=-(1+2\phi)dt^2+2B_idx^idt+[(1-2\psi)\delta_{ij}+2E_{ij}]dx^idx^j, 
\end{equation}
where $E_{ij}$ is symmetric, traceless tensor and the index $i, j$ are raised and lowered by $\delta_{ij}$. We can further decompose $B_i$ and $E_{ij}$ into helicity-0, 1, 2 modes,
\begin{align}
B_i&=\partial_i B+B_i^\text{T} \\
E_{ij}&=\partial_{\langle i}\partial_{j\rangle}E+\partial_{(i}E_{j)}^\text{T}+E_{ij}^{\text{TT}},
\end{align}
where $B$ and $B_i^\text{T}$ are longitudinal and transverse parts of vector $B_i$, $E_i^\text{T}$ is transverse, and $E_{ij}^\text{TT}$ is symmetric, trace-free and transverse, and the angled bracket indices component
\begin{equation}
\partial_{\langle i}\partial_{j\rangle}E=\partial_i\partial_jE-\frac{1}{3}\delta_{ij}\nabla^2E
\end{equation}
is trace-free. By this decomposition, we can separate the action into helicity-0, 1, 2 sectors, since at the quadratic level there is no mixing between different helicities.

\subsection{Helicity-$2$ sector}\label{sec:tensormode}
The second order action of helicity-2 modes is
\begin{align}\label{tensoraction}
S=\frac{M_P^2}{2}\int d^4x~&\beta[(\ddot{E}^\text{TT}_{ij})^2+2\dot{E}^{\text{TT}ij}\nabla^2\dot{E}^\text{TT}_{ij}+(\nabla^2E_{ij}^\text{TT})^2]\nonumber \\
&+(\dot{E}^\text{TT}_{ij})^2+E^{\text{TT}ij}\nabla^2E_{ij}^\text{TT},
\end{align}
which describes two massless helicity-2 degrees of freedom originating from the massless graviton and two massive helicity-2 degrees of freedom coming from the quadratic invariant term $\beta R_{\mu\nu}R^{\mu\nu}$.  Since there is no first class (i.e. gauge) constraint in the helicity-2 modes, there are four helicity-2 degrees of freedom in the theory. Notice that only the $\beta$ term enters in this expression.

Ostrogradski's choice of canonical coordinates is the pair of canonical variables $(E_{ij},\pi_{ij})$ and $(q_{ij},p_{ij})$ defined by
\begin{align}
E_{ij}\equiv E_{ij}^\text{TT}  &\longleftrightarrow \pi^{ij}=2\dot{E}^{\text{TT}ij}+\beta(-2\dddot{E}^{\text{TT}ij}+4\nabla^2\dot{E}^{\text{TT}ij})\nonumber \\
q_{ij}\equiv\dot{E}_{ij}^\text{TT} &\longleftrightarrow p^{ij}=2\beta\ddot{E}^{\text{TT}ij}.
\end{align}
One might notice that in Ostrogradski's formalism the two canonical variables $E_{ij}$, $q_{ij}$ have different dimensionalities, the field $E_{ij}$ is dimensionless while $q_{ij}$ has mass dimension 1 and thus the dimension of canonical momenta are different. The dimensionality is not particularly important -- in principle one can rescale $q_{ij}=M_P^{-1}\dot{E}_{ij}$ to make the two canonical variable at the same footing.

Using the Legendre transform, we construct the Hamiltonian as usual
\begin{align}\label{tensorH}
H=\frac{M_P^2}{2}\int d^3x~& \frac{p^{ij}p_{ij}}{4\beta}+\pi^{ij}q_{ij}-2\beta q^{ij}\nabla^2q_{ij}-q^{ij}q_{ij}\nonumber \\
&-\beta\nabla^2E^{ij}\nabla^2E_{ij}-E^{ij}\nabla^2E_{ij}.
\end{align}
It is easy to check that the Hamiltonian (\ref{tensorH}) generates the equations of motion for the 4 canonical variables via the Poisson Bracket $d(\cdot)/dt{}=[\cdot,H]$. The important point here is that the Hamiltonian is linearly dependent on $\pi^{ij}$ in the second term and hence the Hamiltonian is unbounded from below -- the $\pi^{ij}q_{ij}$ term can be arbitrarily negative when $q_{ij}>0$, $\pi^{ij}\rightarrow -\infty$ or vice versa.

This instability is often called a ``ghost'', i.e. a dynamical degree of freedom with the wrong sign kinetic term.  To see this, we can explicitly diagonalize the Hamiltonian by the following canonical transformation
\begin{align}
\psi_{ij}&=\frac{1}{\sqrt{2}}\left(\frac{p_{ij}}{2}-\beta\nabla^2E_{ij}\right)\nonumber \\
\phi_{ij}&=\frac{1}{\sqrt{2}}\left(-\frac{p_{ij}}{2}+\beta\nabla^2E_{ij}+E_{ij}\right)\nonumber \\
p_{\psi ij} &=\sqrt{2}\left(\pi_{ij}-2\beta\nabla^2q_{ij}-2q_{ij}\right)\nonumber\\\
p_{\phi ij}&=\sqrt{2}\left(\pi_{ij}-2\beta\nabla^2q_{ij}\right),
\end{align}
and the Hamiltonian thus becomes
\begin{align}
H=&\frac{M_P^2}{2}\int d^3x~\frac{p_{\phi ij}p^{ij}_\phi}{2}-\frac{\phi_{ij}\nabla^2\phi^{ij}}{2}\nonumber \\
&-\left(\frac{p_{\psi ij}p^{ij}_\psi}{2}-\frac{\psi_{ij}\nabla^2\psi^{ij}}{2}-\frac{\psi_{ij}\psi^{ij}}{2\beta}\right),
\end{align}
where the $(\psi, p_{\psi})$ pair is ghostlike. In the classical theory, the unboundedness of the Hamiltonian leads to instabilities as the phase space for negative energy higher frequency modes become unbounded below \footnote{This is to be contrasted with \emph{tachyonic} instability -- which is the ``exponential blowing up'' of each individual mode.}. In the quantum theory, while this instability does not prevent us from identifying a vacuum state and then constructing the Fock space of many particle states, imposition of positivity in the energy of all particle states will lead to some states possessing negative norms, i.e. ghosts. One can further excise these unphysical negative norm states from the Fock space, but this generically leads to violations of unitarity.   For a review of the quantization issues with such theories, see Appendix \ref{sect:appendix1}.

\subsection{Helicity-1 sector}\label{sec:vectormode}
The second order action of helicity-1 modes can be written by the gauge invariant variable $v_i=\sqrt{-\nabla^2}(B_i^\text{T}-\dot{E}^\text{T}_i)$ \footnote{One should not be unduly worried by the appearance of the non-local square root of Laplacian operator. Recall that the Laplacian operator $-\nabla^2$ has zero or positive eigenvalues $\lambda_k$, e.g. $-\nabla^2 \phi_k = \lambda_k \phi_k$ with $\lambda \geq 0$. Formally, $\sqrt{-\nabla^2} u = \sum_k c_k \lambda_k^{1/2}\phi_k$ (as long as both $u$ and $\phi_k$ vanish at the boundary), i.e. $u= \sum_k c_k \phi_k$.} 
\begin{equation}\label{vectoraction}
S=\frac{M_P^2}{2}\int d^4x~\frac{\beta}{2}(\dot{v}_i\dot{v}^i+v_i\nabla^2 v^i+\frac{1}{\beta}v_iv^i).
\end{equation}
The action describes a vector with mass $m^{-2}=-\beta$ and the sign of $\beta$ also decides the overall sign of the action, i.e. if $\beta<0$, the helicity-1 modes are ghostlike.  The Euler-Lagrange equation of action (\ref{vectoraction}) is
\begin{equation}\label{vector EL eqn}
\left[\beta\left(\frac{d^2}{dt^2}-\nabla^2\right)-1\right]v_i=0, 
\end{equation}
which can be solved by Fourier transform, and the solutions are harmonic oscillators with frequency $w_\bold{p}^2=\bold{p}^2-\frac{1}{\beta}$. The canonical momentum conjugate to $v_i$ is as usual defined by
\begin{equation}
p_{vi}=\frac{\delta S}{\delta\dot{v}_i}=\beta\dot{v}_i,
\end{equation}
and since we use the gauge invariant variable to write the action, there is no constraint in helicity-1 sector and the Hamiltonian is 
\begin{equation}\label{vectorH}
H=\frac{M_P^2}{2}\int d^3x~\frac{p_{vi}p_v^i}{2\beta}-\frac{\beta}{2}v_i\nabla^2v^i-\frac{1}{2}v_iv^i.
\end{equation}
If we choose $\beta>0$, then $m^2<0$, which means the theory is tachyonic. On the other hand, if we choose $\beta<0$ in eq.(\ref{vectorH}), the Hamiltonian will be negative definite and thus ghostlike. 
One can see that if $\beta<0$,  we can perform a canonical transformation of the variables into ``canonically normalized'' form $\sqrt{-\beta}v_i\rightarrow v_i$, $(-\beta)^{-\frac{1}{2}}p_{vi}\rightarrow p_{vi}$ with the Hamiltonian
\begin{equation}
H=\frac{M_P^2}{2}\int d^3x~-\frac{p_{vi}p_v^i}{2}+\frac{1}{2}v_i\nabla^2v^i+\frac{1}{2\beta}v_iv^i,
\end{equation}
where the mass of the helicity-1 ghost is $m^{-2}=-\beta$. In other words, the helicity-1 modes are either tachyonic $(\beta>0)$ or ghostlike $(\beta<0)$.

\subsection{Helicity-0 sector}\label{sec:scalarmode}
The second order action for helicity-0 modes is more complicated. With the help of two gauge invariant variables
\begin{align}
\Phi&=\phi+\dot{B}-\ddot{E}\nonumber, \\
\Psi&=\psi+\frac{1}{3}\nabla^2E,
\end{align}
the action can be written as
\begin{eqnarray} \label{scalaraction}
S&=&\frac{M_P^2}{2}\int d^4x~(-6\dot{\Psi}^2-2\Psi\nabla^2\Psi+4\Psi\nabla^2\Phi)\nonumber \\
&&+4(\beta+3\alpha)(3\ddot{\Psi}^2+4\dot{\Psi}\nabla^2\dot{\Psi}+2\ddot{\Psi}\nabla^2\Phi) \nonumber \\
&&+2(3\beta+8\alpha)(\nabla^2\Psi)^2+2(\beta+2\alpha)(\nabla^2\Phi)^2 \nonumber \\
&&-4(\beta+4\alpha)\nabla^2\Psi\nabla^2\Phi.
\end{eqnarray}

There are two scalar functions in the action, and because of the second order time derivatives on $\Psi$, there are three naive degrees of freedom \footnote{As in eq.(\ref{eqn:S1}), an extra time derivative in the action will bring you two more dimensions of phase space, i.e., one more degree of freedom.}. One degree of freedom will eventually be removed by gauge constraint and the helicity-0 modes sector are in general consist of two degrees of freedom. Notice that all the second order time derivatives appear on the second line with the coefficient $(\beta+3\alpha)$ -- this is the well-known fact \cite{Stelle:1976gc} that if we choose $\beta+3\alpha=0$, the massive scalar will be frozen and removed from the theory because of its infinite mass. The only degree of freedom in this sector is the helicity-0 mode of massive graviton.

On the other hand, we know that $\beta=0$ is simply an $f(R)$ type theory which is degenerate and hence is also ghost-free -- this fact is not manifest in the eq.(\ref{scalaraction}) above if we simply set $\beta=0$.  However, when $\beta=0$, the action can be rearranged as
\begin{align} \label{frscalaraction}
S=\frac{M_P^2}{2}\int d^4x~&(-6\dot{\Psi}^2-2\Psi\nabla^2\Psi+4\Psi\nabla^2\Phi)\nonumber \\
&+4\alpha(3\ddot{\Psi}-2\nabla^2\Psi+\nabla^2\Phi)^2,
\end{align}
where we have suggestively written the second line in the action (\ref{frscalaraction}) as a complete square. By varying $\Phi$, we obtain
\begin{equation} \label{eqn:eqnbeta0}
\nabla^2 \Phi=\left(-\frac{1}{2\alpha}+2\nabla^2\right)\Psi-3\ddot{\Psi}.
\end{equation}
Inserting eq.(\ref{eqn:eqnbeta0}) back into the action (\ref{frscalaraction}) we obtain the action of a single non-ghostlike massive scalar field
\begin{align} 
S=\frac{M_P^2}{2}\int d^4 x~&6\dot{\Psi}^2+6\Psi\nabla^2\Psi-\frac{2}{\alpha}\Psi^2
\end{align}
as we would expect for $f(R)$ type theories. It is clear that since the action is only dependent on $\Psi$ and $\dot{\Psi}$, there is only one ghost-free d.o.f.. Notice that if $\alpha<0$ this scalar is a tachyonic unstable d.o.f., which is consistent with the general $f(R)$ gravity, where we require $f''(R)>0$ to avoid tachyonic instability \cite{Starobinsky:2007hu, DeFelice:2010aj}. By setting $\alpha\rightarrow 0$ means that the mass term blows up and rendering this d.o.f. non-dynamical, i.e. it reduces to simple General Relativity.  

Harking back to the action for general $\alpha$ and $\beta$, eq.(\ref{scalaraction}), Ostrogradski's choice of canonical coordinates is
\begin{align}
\Phi\equiv \Phi&\longleftrightarrow p_\Phi\equiv0\nonumber \\
\Psi\equiv\Psi&\longleftrightarrow p_\Psi \equiv \frac{\delta S}{\delta \dot{\Psi}}\nonumber \\
\chi\equiv \dot{\Psi}&\longleftrightarrow p_\chi\equiv 8(\beta+3\alpha)(3\ddot{\Psi}+\nabla^2\Phi),
\end{align}
where the choice $\beta+3\alpha=0$ means that $p_\chi=0$ becomes a \emph{primary} constraint instead of an additional d.o.f..

The Hamiltonian can be expressed by the canonical coordinates
\begin{align}
H=&\frac{M_P^2}{2}\int d^3x ~ p_\Psi \chi+\frac{p_\chi^2}{48(\beta+3\alpha)}-\frac{p_\chi\nabla^2\Phi}{3}\nonumber \\
&+(6\chi^2+2\Psi\nabla^2\Psi-4\Psi\nabla^2\Phi)\nonumber \\
&-16(\beta+3\alpha)\chi\nabla^2\chi-2(3\beta+8\alpha)(\nabla^2\Psi)^2\nonumber\\
&+4(\beta+4\alpha)\nabla^2\Psi\nabla^2\Phi-\frac{2\beta}{3}(\nabla^2\Phi)^2.
\end{align}
The primary constraint is $\varphi_1: p_\Phi=0$ and all the constraints can be generated by the consistency relation
\begin{equation}\label{varphi2}
\varphi_2: \nabla^2\left(\frac{p_\chi}{3}+4\Psi-4(\beta+4\alpha)\nabla^2\Psi+\frac{4\beta}{3}\nabla^2\Phi\right)\approx 0,
\end{equation}
where $\approx$ means ``weak equality'' (i.e. only satisfied when the variables are on-shell) -- see \cite{Henneaux} for a discussion on this point.

Since $\varphi_1$, $\varphi_2$ are second class \footnote{Note that in the case of $\beta=0$, the constraints $\varphi_1$ and $\varphi_2$ are not second class and the theory will contain two more constraints. The reduced phase space is then two dimensional and the Hamiltonian is bounded below if $\alpha>0$, same as the conclusion of full $f(R)$ theory.}, we can use them to reduce the phase space $(\Phi, p_\Phi)$, and the reduced Hamiltonian is 
\begin{align}\label{scalarH}
H_R=&\frac{M_P^2}{2}\int d^3x~  p_\Psi \chi+\frac{1}{\beta}p_\chi[1-(\beta+4\alpha)\nabla^2]\Psi\nonumber \\
&+\frac{(\beta+2\alpha)}{16\beta(\beta+3\alpha)}p_\chi^2+6\chi^2-16(\beta+3\alpha)\chi\nabla^2\chi\nonumber \\
&+\frac{6}{\beta}\Psi^2-(10+\frac{48\alpha}{\beta})\Psi\nabla^2\Psi+\frac{32\alpha(\beta+3\alpha)}{\beta}(\nabla^2\Psi)^2.
\end{align} 
The linear dependence of $p_\Psi$ again renders the Hamiltonian unbounded from below.

In order to see the mass content of helicity-0 modes, we will need to further diagonalize the Hamiltonian by the following canonical transformation:
\begin{align}
Q_1&=\sqrt{3}\left(\frac{p_\chi}{6}-\frac{8(\beta+3\alpha)}{3}\nabla^2\Psi+2\Psi\right)\nonumber \\
Q_2&=\sqrt{3}\left(\frac{p_\chi}{6}-\frac{8(\beta+3\alpha)}{3}\nabla^2\Psi\right)\nonumber \\
P_1&=\frac{1}{\sqrt{3}}\left(\frac{p_\Psi}{2}-8(\beta+3\alpha)\nabla^2 \chi\right)\nonumber\\
P_2&=\frac{1}{\sqrt{3}}\left(-\frac{p_\Psi}{2}+8(\beta+3\alpha)\nabla^2 \chi-6\chi\right).
\end{align}
The diagonalized Hamiltonian is then
\begin{align}
H_R=\frac{M_P^2}{2}\int d^3x~&-\frac{P_1^2}{2}+\frac{1}{2}Q_1\nabla^2Q_1+\frac{1}{2\beta}Q_1^2 \nonumber \\
&+\frac{P_2^2}{2}-\frac{1}{2}Q_2\nabla^2Q_2+\frac{1}{2}\frac{1}{2(\beta+3\alpha)}Q_2^2.
\end{align}
The reduced Hamiltonian of helicity-0 sector contains two massive degrees of freedom. One is a massive ghost comes from massive graviton with mass $m_1^{-2}=-\beta$ and the other is massive scalar with positive definite kinetic energy, with mass $m_2^{-2}=2(\beta+3\alpha)$.

Let us combine the result from all sectors. In \ref{sec:tensormode}, we saw that there are four helicity-2 degrees of freedom and two of them suffer from ghost-like instabilities. In \ref{sec:vectormode}, the two helicity-1 degrees of freedom are either ghostlike or tachyonic, depending on the sign of $\beta$. In \ref{sec:scalarmode}, one of two scalar degrees of freedom is ghostlike. With $\beta<0$, one can see the unstable modes in helicity-0, 1, 2 sectors are massive with mass $m^{-2}=-\beta$, which corresponds to the massive graviton. This result is derived by the Stelle in his seminal work on higher derivative gravity \cite{Stelle:1977ry} using an auxiliary field methodology. Here we rederived the results using the usual Hamiltonian formalism.

There are two special choices of parameters in the linearized theory. With $\alpha\neq 0, \beta=0$, the massive graviton sector gains an infinite mass and hence becomes non-dynamical. In this case the theory consists of one massless graviton with one massive scalar field (i.e. an  $f(R)$ theory). On the other hand, by taking the limit $\beta+3\alpha=0$, the massive scalar field becomes infinitely massive and hence non-dynamical. In this case, the theory's particle content reduces to  one massive and one massless graviton. With the latter choice and a total minus sign, at the linear level one can have a theory with a healthy massive graviton \cite{Park:2012ds}, since this choice is consistent with the Fierz-Pauli tuning. However, one should expect that the Boulware-Deser ghost \cite{Boulware:1973my} will enter at the nonlinear level.

\section{Stabilization by constraints in Minkowski background} \label{sect::exorcise}
In this section, we will demonstrate how to remove the unstable degrees of freedom by introducing constraints via auxiliary fields. As shown in \cite{Chen:2012au}, this will result in the effective dimensionality of phase space being reduced. Roughly speaking, we impose the constraints such that the auxiliary fields are related to second order time derivative of the unstable fields, resulting in the final equations of motion being second order in time derivatives yet up to fourth order in spatial derivatives. The advantage of preserving spatial part of the ``higher derivative'' component  is that we retain the improved renormalization properties of such theories, at the price of giving up Lorentz invariance. 

One might ask what if we remove the instabilities without explicitly breaking Lorentz invariance? Here we emphasize that we can equally insert constraints to remove the higher spatial derivatives, with the end result being a stable 2nd order theory both in space and time derivatives. For example, the unconstrained helicity-2 action (\ref{tensoraction}) can be written as
\begin{align}
S=\frac{M_P^2}{2}\int d^4 x ~&\beta(\Box E^{TT}_{ij})^2+E^{\text{TT}ij}\Box E_{ij}^\text{TT}. \nonumber
\end{align}
Without the full theory, we do not know how to introduce $\lambda$ into the action without breaking Lorentz invariance while removing the highest time derivative in the equations of motion. The best thing we can do is to couple $\lambda_{ij}$ with $\Box E_{ij}$, and the Lorentz invariance is not explicitly broken by extra terms. We can modify the action as
\begin{align}
S=\frac{M_P^2}{2}\int d^4 x ~&\beta(\Box E^{TT}_{ij}-\lambda_{ij})^2+E^{\text{TT}ij}(\Box E_{ij}^\text{TT}-a \lambda_{ij}). \nonumber
\end{align}
If $a=1$, we force $\lambda_{ij}$ coupling to every $\Box E_{ij}$ and if $a=0$ we only force $\lambda_{ij}$ coupling to those $\Box E_{ij}$ where the $\Box$ cannot be removed by an integration by part. The equations of motion of the theory are
\begin{align}
\delta\lambda:& 2\beta(\Box E_{ij}^{TT}-\lambda_{ij}^{TT})+a E_{ij}^{TT}=0\nonumber \\
\delta E:& 2\beta \Box(\Box E_{ij}^{TT}-\lambda_{ij})+(\Box E_{ij}^{TT}-a \lambda_{ij}^{TT})+\Box E_{ij}^{TT}=0\nonumber, 
\end{align}
which can be written as a single equation of $E_{ij}$
\begin{equation}\nonumber
2(1-a)\Box E_{ij}^{TT}-\frac{a^2}{2\beta}E_{ij}^{TT}=0.
\end{equation}
The equation is either trivial if $a=1$ or a Klein-Gordon equation with mass $m^2=a^2/4\beta(1-a)$ if $a\neq 1$. In both cases, the equations of motion will have same order of time derivatives and spatial derivatives and the improved renormalization properties will not be retained.

For notational simplicity, from now on we drop the traceless notation $B^{T}_i$, $E^{\text{TT}}$, which should be clear from the context.

\subsection{Helicity-2 sector}
We begin by introducing a helicity-2 auxiliary tensor field $\lambda_{ij}$ into the action (\ref{tensoraction}) 
\begin{align}\label{constrainedtensoraction}
S=\frac{M_P^2}{2}\int d^4x~&\beta[(\ddot{E}_{ij}-\lambda_{ij})^2+2\dot{E}^{ij}\nabla^2\dot{E}_{ij}+(\nabla^2E_{ij})^2]\nonumber \\
&+\dot{E}^{ij}\dot{E}_{ij}+E^{ij}\nabla^2E_{ij}+4\beta\lambda^{ij}\nabla^2E_{ij},
\end{align}
where $\lambda_{ij}$ is transverse traceless, which also explicitly breaks Lorentz invariance. The canonical coordinates are
\begin{align}
E_{ij}\equiv E_{ij}  &\longleftrightarrow \pi^{ij}=2\dot{E}^{ij}+\beta(-2\dddot{E}^{ij}+2\dot{\lambda}^{ij}+4\nabla^2\dot{E}^{ij})\nonumber \\
q_{ij}\equiv\dot{E}_{ij} &\longleftrightarrow p^{ij}=2\beta(\ddot{E}^{ij}-\lambda^{ij})\nonumber \\
\lambda_{ij}\equiv\lambda_{ij}&\longleftrightarrow p^{ij}_\lambda=0,
\end{align}
and the Hamiltonian is
 \begin{align}
H=\frac{M_P^2}{2}\int &d^3 x ~\pi^{ij}q_{ij}+\frac{1}{4\beta}p^{ij}p_{ij}-E^{ij}(\beta\nabla^2\nabla^2+\nabla^2)E_{ij}\nonumber \\
&-q^{ij}(1+2\beta\nabla^2)q_{ij}+\lambda^{ij}(p_{ij}-4\beta\nabla^2E_{ij}).
\end{align}
The Poisson bracket of a pair of transverse traceless canonical coordinates can be found as 
\begin{equation}
[E_{ij}(\bold{x}),\pi_{kl}(\bold{y})]_{PB}=\hat{\Lambda}_{ij,kl}\delta^{(3)}(\bold{x}-\bold{y}),
\end{equation}
where $\hat{\Lambda}_{ij,kl}$ is the transverse traceless projection operator defined by $\hat{\Lambda}_{ij,kl}\equiv1/2(\hat{\theta}_{ik}\hat{\theta}_{jl}+\hat{\theta}_{il}\hat{\theta}_{jk}-\hat{\theta}_{ij}\hat{\theta}_{kl})$, while $\hat{\theta}_{ij}\equiv \delta_{ij}-\frac{\partial_i\partial_j}{\partial^2}$ is the transverse projection operator. Since the equations of motion in the Hamiltonian picture are generated by Poisson bracket, the projection operator will preserve the transverse traceless characteristic.

It is clear that $p_{\lambda ij}=0$ is a primary constraint as it is an auxiliary field. Via the consistency relation, we can generate further (traceless and transverse) secondary constraints as follow
\begin{align}
\varphi_1:&p_{\lambda ij}=0,\nonumber \\
\varphi_2:&p_{ij}-4\beta\nabla^2E_{ij}\approx0,\nonumber \\
\varphi_3:&\pi_{ij}-2q_{ij}\approx0,\nonumber \\
\varphi_4:&2(\beta\nabla^2\nabla^2+\nabla^2)E_{ij}-\frac{1}{\beta}p_{ij}\nonumber \\
&+2(-1+2\beta\nabla^2)\lambda_{ij}\approx 0.
\end{align}
We can use the constraints $\varphi_1$, $\varphi_4$ to eliminate the degree of freedom $(\lambda, p_\lambda)$, and use $\varphi_2$, $\varphi_3$ to eliminate $(q, p)$. The coefficients in the action (\ref{constrainedtensoraction}) are chosen such that there are at least four constraints in the theory and there is no $\nabla^2$ in $\varphi_3$ which will generate nonlocal terms in the reduced Hamiltonian.

Using the constraints,  $(q_{ij}, p_{ij})$ can be written as follow
\begin{align}
q_{ij}&=\frac{\pi_{ij}}{2},\nonumber \\
p_{ij}&=4\beta\nabla^2E_{ij},
\end{align}
and the reduced Hamiltonian becomes 
 \begin{align}\label{reducedH}
H_R=&\frac{M_P^2}{2}\int d^3 x~\frac{1}{4}\pi^{ij}(1-2\beta\nabla^2)\pi_{ij}\nonumber \\
&+E^{ij}(-\nabla^2+3\beta\nabla^2\nabla^2)E_{ij}.
\end{align}
To check whether the reduced Hamiltonian is bounded from below, we will explicitly quantize the theory. Similar to QED in the Coulomb gauge, one can follow Dirac's method to quantize the constrained system. We first write down the generalized version of Poisson bracket (i.e. Dirac bracket), which generates time evolution of any fields in constrained theory while preserving all the constraints. We then promote all the fields to operators and the commutators of two fields now become $i$ times the Dirac bracket of them.

To write down the Dirac bracket, we first define a matrix $C_{ab}\equiv [\varphi_a,\varphi_b]_{PB}$, 
\[
 C_{ab;ij,kl} (\bold{x},\bold{y})= \begin{bmatrix}
       0&0&0&-\hat{a}\\
0&0&-\hat{a}&0\\
0&\hat{a}&0&-\hat{b}\\
\hat{a}&0&\hat{b}&0
     \end{bmatrix}\hat{\Lambda}_{ij,kl}\delta^{(3)}(\bold{x-y}),
\]
where $\hat{a}$ and $\hat{b}$ are two operators $\hat{a}\equiv2(-1+2\beta\nabla^2)$ and $\hat{b}\equiv2(\beta\nabla^2\nabla^2+\nabla^2-1/\beta)$.
The inverse of $C_{ab}$ is 
\[
 C^{-1;ab;ij,kl}= \begin{bmatrix}
       0&\hat{a}^{-2}\hat{b}&0&\hat{a}^{-1}\\
-\hat{a}^{-2}\hat{b}&0&\hat{a}^{-1}&0\\
0&-\hat{a}^{-1}&0&0\\
-\hat{a}^{-1}&0&0&0
     \end{bmatrix}\hat{\Lambda}^{ij,kl}\delta^{(3)}(\bold{x-y}),
\]
and the Dirac bracket of two field $X$, $Y$ is defined by 
\begin{equation}
[X, Y]_D=[X,Y]_{PB}-[X,\Phi_{a,ij}]_{PB} C^{-1;ab;ij,kl}[\Phi_{b,kl}, Y].\nonumber\\
\end{equation}
Equipped with Dirac bracket, one can use the reduced Hamiltonian to write down the equations of motion of this system
\begin{align}\label{eomofe}
\dot{E}_{ij}&=[E_{ij}, H_R]_D=\frac{1}{2}\pi_{ij}\\
\dot{\pi}_{ij}&=\frac{(-2\nabla^2+6\beta\nabla^2\nabla^2)}{(-1+2\beta\nabla^2)}E_{ij}.
\label{eomofpi}
\end{align}
Using eq.(\ref{eomofe}) and  eq.(\ref{eomofpi}), we find  
\begin{equation}\label{ELequation}
\ddot{E}_{ij}=\frac{(-\nabla^2+3\beta\nabla^2\nabla^2)}{(-1+2\beta\nabla^2)}E_{ij},
\end{equation}
which is the Euler-Lagrange equation of the action (\ref{constrainedtensoraction}). We can solve eq.(\ref{ELequation}) by taking the Fourier transform
\begin{equation}
E_{ij}(\bold{x}, t)=\int\frac{d^3p}{(2\pi)^3}~e^{i\bold{p}\cdot\bold{x}}\tilde{E}_{ij}(\bold{p},t),
\end{equation}
where $\tilde{E}_{ij}(\bold{p},t)$ satisfies
\begin{equation}
\left[\frac{d^2}{dt^2}-\frac{(\bold{p}^2+3\beta\bold{p}^4)}{(-1-2\beta\bold{p}^2)}\right]\tilde{E}_{ij}(\bold{p},t)=0.
\end{equation}
For any $\bold{p}$, $\tilde{E}_{ij}(\bold{p},t)$ is a harmonic oscillator with frequency $w_\bold{p}=\sqrt{\frac{(\bold{p}^2+3\beta\bold{p}^4)}{(1+2\beta\bold{p}^2)}}$, where $w_\bold{p}^2$ is positive definite if $\beta>0$.

In order to quantize the theory, we write $E_{ij}$, $\pi_{ij}$ as linear summation of creation and annihilation operators $a^{r\dagger}_{\bold{p}}$, $a^r_\bold{p}$,
\begin{align}
E_{ij}(\bold{x})&=\int \frac{d^3p}{(2\pi)^3}~\frac{1}{\sqrt{2|w_\bold{p}|}}\sum^2_{r=1}\epsilon^r_{ij}(\bold{p})(a^r_\bold{p}e^{i\bold{p}\cdot\bold{x}}+a^{r\dagger}_{\bold{p}}e^{-i\bold{p}\cdot\bold{x}})\nonumber \\
\pi_{kl}(\bold{x})&=\int \frac{d^3p}{(2\pi)^3}~-2i\sqrt{\frac{|w_\bold{p}|}{2}}\sum^2_{r=1}\epsilon^r_{kl}(\bold{p})(a^r_\bold{p}e^{i\bold{p}\cdot\bold{x}}-a^{r\dagger}_{\bold{p}}e^{-i\bold{p}\cdot\bold{x}}),
\end{align} 
where the coefficients are chosen in such a way that they solve the equations of motion eqs.(\ref{eomofe}), (\ref{eomofpi}), and the superscript $r$ labels the polarizations. The symmetric transverse traceless tensor $\epsilon_{ij}^r$  satisfies $p^i\epsilon_{ij}^r=\delta^{ij}\epsilon_{ij}^r=0$, and is normalized as
\begin{equation}\label{normalization}
\sum_{i,j}\epsilon_{ij}^r(\bold{p})\epsilon^{s,ij}(\bold{p})=\frac{\delta^{rs}}{2(1+2\beta \bold{p^2})},
\end{equation}
with the completeness relation
\begin{equation}\label{complete relation}
\sum_{r=1}^2\epsilon_{ij}^r(\bold{p})\epsilon^{r}_{kl}(\bold{p})=\frac{\Lambda_{ij,kl}(\bold{p})}{2(1+2\beta \bold{p}^2)}.
\end{equation}
The operator $\Lambda_{ij,kl}(\bold{p})$ is defined by replacing all the $\nabla^2$ in the transverse traceless projection operator by $-\bold{p}^2$. One can calculate the Dirac bracket of  ($E_{ij}$, $\pi_{kl}$) and the commutator of the two operators is thus
\begin{equation}\label{commutator1}
[E_{ij}(\bold{x}),\pi_{kl}(\bold{y})]=-i\Lambda_{ij,kl}\left[\frac{1}{(-1+2\beta\nabla^2)}\right]\delta^{(3)}(\bold{x-y}).
\end{equation}
With the normalization eq.(\ref{normalization}) and the completeness relation eq.(\ref{complete relation}), the commutation relation eq.(\ref{commutator1}) is equivalent to
\begin{align}
[a^r_\bold{p},a^{s}_\bold{q}]&=[a^{r, \dagger}_\bold{p},a^{s, \dagger}_\bold{q}]=0, \nonumber \\
[a^{r}_\bold{p},a^{s, \dagger}_\bold{q}]&=(2\pi^3)\delta^{rs}\delta^{(3)}(\bold{p}-\bold{q}).
\end{align}
One can thus rewrite the reduced Hamiltonian (\ref{reducedH}) as creation and annihilation operators
\begin{equation}
H_R=\frac{M_P^2}{2}\int\frac{d^3p}{(2\pi)^3}~|w_p|\left(\sum_{r=1}^2a^{r,\dagger}_\bold{p}a^r_\bold{p}+\frac{1}{2}(2\pi)^3\delta^{rr}\delta^{(3)}(0)\right).
\end{equation}
The energy spectrum is real and bounded from below if $w_p^2$ is positive definite as long as $\beta>0$.

\subsection{Helicity-1 sector}
We now turn to the attention of the helicity-1 unstable modes. As shown in \ref{sec:vectormode}, this sector is tachyonic if $\beta<0$ and ghostlike if $\beta>0$. As usual, we will remove it by modifying the action (\ref{vectoraction}) with the introduction of a helicity-1 field $\lambda_i$
\begin{equation}\label{mdvectoraction}
S=\frac{M_P^2}{2}\int d^4 x ~\frac{\beta}{2}\left[(\dot{v}_i-\lambda_i)^2+v_i\nabla^2 v^i+\frac{1}{\beta}v_iv^i\right].
\end{equation}
Ostrogradski's choice of canonical coordinates is
\begin{align}
v_i\equiv v_i&\longleftrightarrow p_{v}^i=\beta(\dot{v}^i-\lambda^i)\nonumber\\
\lambda_i\equiv \lambda_i&\longleftrightarrow p_{\lambda}^i=0,
\end{align}
and the Hamiltonian is 
\begin{align}
H=\frac{M_P^2}{2}\int d^3 x~ \frac{p_v^ip_{vi}}{2\beta}+p_v^i\lambda_i-\frac{\beta}{2}v_i\nabla^2v^i-\frac{1}{2}v_iv^i.
\end{align}
There are four constraints in the theory, which can be found as
\begin{align}
\varphi_1&: p^i_\lambda=0\nonumber \\
\varphi_2&: p^i_v\approx0\nonumber \\
\varphi_3&:v^i+\beta\nabla^2v^i\approx 0\nonumber \\
\varphi_4&:\frac{p^i_v}{\beta}+\nabla^2p^i_v+\lambda^i+\beta\nabla^2\lambda^i\approx 0.
\end{align}
If we use the four constraints to eliminate $(v_i, p^i_v)$, $(\lambda_i, p_\lambda^i)$, the physical phase space will be zero dimensional and the reduced Hamiltonian vanishes.

\subsection{Helicity-0 sector}
Finally we introduce a helicity-0 field $\lambda$ into the action(\ref{scalaraction})
\begin{align}\label{constrainedscalaraction}
S=&\frac{M_P^2}{2}\int d^4 x ~(-6\dot{\Psi}^2-2\Psi\nabla^2\Psi+4\Psi\nabla^2\Phi)\nonumber \\
&+4(\beta+3\alpha)(3\ddot{\Psi}^2+4\dot{\Psi}\nabla^2\dot{\Psi}+2\ddot{\Psi}\nabla^2\Phi) \nonumber \\
&+2(3\beta+8\alpha)(\nabla^2\Psi)^2+2(\beta+2\alpha)(\nabla^2\Phi)^2 \nonumber \\
&-4(\beta+4\alpha)\nabla^2\Psi\nabla^2\Phi+32(\beta+3\alpha)\lambda\nabla^2\Psi\nonumber \\
&+12(\beta+3\alpha)(\lambda^2-2\ddot{\Psi}\lambda-\frac{2}{3}\lambda\nabla^2\Phi)+A\lambda \Psi,
\end{align}
where $A$ is some arbitrary real constant. There are four naive degrees of freedom, whose canonical variables are  
\begin{align}
\Phi\equiv \Phi &\longrightarrow p_\Phi=0\nonumber \\
\Psi\equiv \Psi &\longrightarrow p_\Psi=\frac{\delta S}{\delta \dot{\Psi}}\nonumber \\
\chi\equiv \chi &\longrightarrow p_\chi=8(\beta+3\alpha)[3(\ddot{\Psi}-\lambda)+\nabla^2\Phi]\nonumber \\
\lambda\equiv \lambda &\longrightarrow p_\lambda=0.
\end{align}
The Hamiltonian can be written as
\begin{align}
H=&\frac{M_P^2}{2}\int d^3x  ~p_\Psi \chi+\frac{p_\chi^2}{48(\beta+3\alpha)}-\frac{p_\chi\nabla^2\Phi}{3}\nonumber \\
&+\left(6\chi^2+2\Psi\nabla^2\Psi-4\Psi\nabla^2\Phi\right)\nonumber \\
&-16(\beta+3\alpha)\chi\nabla^2\chi-2(3\beta+8\alpha)(\nabla^2\Psi)^2\nonumber\\
&+4(\beta+4\alpha)\nabla^2\Psi\nabla^2\Phi-\frac{2\beta}{3}(\nabla^2\Phi)^2\nonumber \\
&+\lambda(p_\chi-A\Psi)-32(\beta+3\alpha)\lambda\nabla^2\Psi,
\end{align}
where the constraints of this theory are
\begin{align}
\varphi_1&: p_\Phi=0\nonumber \\
\varphi_2&: p_\lambda=0\nonumber \\
\varphi_3&:p_\chi-A\Psi-32(\beta+3\alpha)\nabla^2\Psi\approx 0\nonumber \\
\varphi_4&:\nabla^2\left[\frac{p_\chi}{3}+4\Psi-4(\beta+4\alpha)\nabla^2\Psi+\frac{4\beta}{3}\nabla^2\Phi\right]\approx 0\nonumber \\
\varphi_5&:p_\Psi+(12+A)\chi\approx0 \nonumber \\
\varphi_6&:\frac{(12+A)p_\chi}{24(\beta+3\alpha)}+2(6+A)\lambda+32(\beta+3\alpha)\nabla^2\lambda\nonumber \\
&+4(3\beta+8\alpha)\nabla^2\nabla^2\Psi-4(\beta+4\alpha)\nabla^2\nabla^2\Phi\nonumber \\
&-4\nabla^2\Psi-\frac{A}{3}\nabla^2\Phi\approx 0.
\end{align}
We use the six constraints to eliminate \emph{three} pairs of canonical coordinates $(\Phi, p_\Phi)$, $(\lambda, p_\lambda)$, and $(\chi, p_\chi)$, reducing the Hamiltonian to become
\begin{align}
H_R=\frac{M_P^2}{2}\int d^3x  ~&\frac{-p_\Psi}{(12+A)^2}[(6+A)+16(\beta+3\alpha)\nabla^2]p_\Psi\nonumber \\
&+\frac{1}{\beta}\left[(6+A)+\frac{A^2(\beta+2\alpha)}{16(\beta+3\alpha)}\right]\Psi^2\nonumber \\
&+\left[(22+\frac{48\alpha}{\beta})+\frac{A}{\beta}(3\beta+4\alpha)\right]\Psi\nabla^2\Psi\nonumber \\
&+\frac{32(\beta+3\alpha)(\beta+\alpha)}{\beta}(\nabla^2\Psi)^2,
\end{align} 
which generates the evolution of a single dynamical variable $\Psi$. It is clear that it can be made positive definite with some parameter choice, for example, ($A=-8, \alpha=0, \beta>0$). We can check that the quantum theory is also stable in the following manner. First we find the Dirac bracket of the theory as usual, which can be used to find the equations of motion 
\begin{align}
\dot{\Psi}=&-\frac{1}{(12+A)}p_\Psi \label{psieom}\\
\dot{p}_\Psi=&-\frac{(12+A)}{2(6+A)+32(\beta+3\alpha)\nabla^2}\times\nonumber \\
&\left\{\frac{2}{\beta}\left[(6+A)+\frac{A^2(\beta+2\alpha)}{16(\beta+3\alpha)}\right]\Psi\right.\nonumber \\
&+2\left[(22+\frac{48\alpha}{\beta})+\frac{A}{\beta}(3\beta+4\alpha)\right]\nabla^2\Psi\nonumber \\
&\left.+\frac{64(\beta+3\alpha)(\beta+\alpha)}{\beta}\nabla^2\nabla^2\Psi\right\}.\label{eq62}
\end{align}
One can check that eqs. (\ref{psieom}), (\ref{eq62}) reproduce the Euler-Lagrange equation. We solve these equations by taking the Fourier transform
\begin{equation}
\Psi(\bold{x},t)=\int\frac{d^3p}{(2\pi)^3}~e^{i\bold{p}\cdot\bold{x}}\tilde{\Psi}(\bold{p},t)
\end{equation}
as usual, and the solution $\tilde{\Psi}(\bold{p},t)$ is a harmonic oscillator with frequency
\begin{align}
w^2_{\bold{p}}=&\frac{1}{\beta[16(\beta+3\alpha)\bold{p}^2-(6+A)]}\left\{\frac{{}^{}}{}32(\beta+3\alpha)(\beta+\alpha)\bold{p}^4\right.\nonumber \\
&\left.-[\beta(3A+22)+\alpha(4A+48)]\bold{p}^2\right.\nonumber \\
&\left.+\left[6+A+\frac{A^2(\beta+2\alpha)}{16(\beta+3\alpha)}\right]\right\}.
\end{align}
Following the usual quantization rules, we define the commutator of $\Psi$ and $p_\Psi$ to be ``$i$" times their classical Dirac bracket, i.e.  
\begin{equation}\label{commutator}
[\Psi, p_\Psi]\equiv\frac{(12+A)i}{2(6+A)+32(\beta+3\alpha)\nabla^2}\delta^{(3)}(\bold{x}-\bold{y}).
\end{equation} 
By expanding $\Psi$, $p_\Psi$ as linear summation of creation and annihilation operators $a_{\bold{p}}^{\dagger}$, $a_\bold{p}$
\begin{align}
\Psi&=\int\frac{d^3p}{(2\pi)^3}~\frac{1}{\sqrt{2w_\bold{p}}}\frac{[a_\bold{p}e^{i\bold{p}\cdot\bold{x}}+a_\bold{p}^\dagger e^{i\bold{p}\cdot\bold{x}}]}{\sqrt{[32(\beta+3\alpha)\bold{p}^2-2(6+A)]}}\\
p_\Psi&=\int\frac{d^3p}{(2\pi)^3}~\sqrt{\frac{w_\bold{p}}{2}}\frac{i(12+A)[a_\bold{p}e^{i\bold{p}\cdot\bold{x}}-a_\bold{p}^\dagger e^{i\bold{p}\cdot\bold{x}}]}{\sqrt{[32(\beta+3\alpha)\bold{p}^2-2(6+A)]}},
\end{align}
we obtain the usual result that the commutator eq.({\ref{commutator}}) is consistent with the Fock space commutators 
\begin{align}
[a_\bold{p},a_\bold{q}]&=[a^{, \dagger}_\bold{p},a^{\dagger}_\bold{q}]=0, \nonumber \\
[a_\bold{p},a^{\dagger}_\bold{q}]&=(2\pi^3)\delta^{(3)}(\bold{p}-\bold{q}).
\end{align}
Using these operators, the reduced Hamiltonian can then be written as
\begin{equation}
H_R=\frac{M_P^2}{2}\int\frac{d^3p}{(2\pi)^3}~|w_\bold{p}|\left(\sum_{r=1}^2a^{\dagger}_\bold{p}a_\bold{p}+\frac{1}{2}(2\pi)^3\delta^{(3)}(0)\right),
\end{equation}
which is bounded from below as long $w_\bold{p}^2$ is positive definite, a condition which is satisfied by suitable choices of the parameters $A,\alpha$ and $\beta$.

\section{Quadratic action around de Sitter background}\label{sect::dS}
In this and next section, we will show how to remove the unstable degrees of freedom in each helicity sector in de Sitter background. In order to separate the action into helicity-0, 1, 2 sectors, we first parameterize the metric fluctuation as 
\begin{align}\label{metricds}
ds^2=a^2(t)&\left[-(1+2\phi)dt^2+2B_idx^idt\right.\nonumber\\
&\left.+[(1-2\psi)\delta_{ij}+2E_{ij}]dx^idx^j\right], 
\end{align}
where $t$ is conformal time in this and next section, and in de Sitter background we have $a(t)=-\frac{1}{Ht}$, $\Lambda=3H^2$. We can again decompose $B_i$, $E_{ij}$ into 
\begin{align}
B_i&=\partial_i B+B_i^\text{T} \\
E_{ij}&=\partial_{\langle i}\partial_{j\rangle}E+\partial_{(i}E_{j)}^\text{T}+E_{ij}^{\text{TT}}.
\end{align}

\subsection{Helicity-2 sector}\label{sec:tensormodedS}
The second order action of helicity-2 modes in de Sitter background is 
\begin{align}\label{tensoractiondS}
S=\frac{M_P^2}{2}&\int d^4 x ~\beta[(\ddot{E}^\text{TT}_{ij})^2+2\dot{E}^{\text{TT}ij}\nabla^2\dot{E}^\text{TT}_{ij}+(\nabla^2E_{ij}^\text{TT})^2]\nonumber \\
&+c a^2(t)[(\dot{E}^\text{TT}_{ij})^2+E^{\text{TT}ij}\nabla^2E_{ij}^\text{TT}],
\end{align}
which reduces to the unconstrained Minkowski case if $a(t)\rightarrow 1$, and $H^2\rightarrow 0$. We have also defined the dimensionless parameter
\begin{equation} \label{cdefine}
c\equiv 1+8H^2(\beta+3\alpha).
\end{equation}
From eq.(\ref{fullaction}), we note that in de Sitter space $R\approx H^2$, and hence we generally expect that $|\alpha| H^2 \ll 1$ and $|\beta| H^2 \ll 1$ if we reasonably suppose that the higher derivative terms are corrections to usual General Relativity. This means that generically we should expect $c>0$ unless the higher derivative terms dominate. As an aside, note that in the special case $\beta+3\alpha=0$, $c=1$.

Since there is no constraint in the helicity-2 modes, there are four helicity-2 degrees of freedom in the theory. Ostrogradski's choice of canonical coordinates is
\begin{align}
E_{ij}\equiv E_{ij}^\text{TT}  &\longleftrightarrow \pi^{ij}\nonumber \\
q_{ij}\equiv\dot{E}_{ij}^\text{TT} &\longleftrightarrow p^{ij}=2\beta\ddot{E}^{\text{TT}ij},
\end{align}
and the Hamiltonian is
\begin{align}\label{tensorHdS}
H=&\frac{M_P^2}{2}\int d^3x~ \frac{p^{ij}p_{ij}}{4\beta}+\pi^{ij}q_{ij}-2\beta q^{ij}\nabla^2q_{ij}\nonumber \\
&-ca^2(t)q^{ij}q_{ij}-\beta\nabla^2E^{ij}\nabla^2E_{ij}-ca^2(t)E^{ij}\nabla^2E_{ij}.
\end{align}
As in eq.(\ref{tensorHdS}), the linear dependence of $\pi^{ij}$ is the signal of Ostrogradski's instability. The $\pi^{ij}q_{ij}$ term can be arbitrarily negative when $q_{ij}>0$, $\pi^{ij}\rightarrow -\infty$ or vice versa and hence the Hamiltonian is thus unbounded from below.

\subsection{Helicity-1 sector}\label{sec:vectormodedS}
The action up to quadratic level of helicity-1 modes can be written using the gauge invariant variable $v_i=\sqrt{-\nabla^2}(B_i^\text{T}-\dot{E}^\text{T}_i)$
\begin{equation}\label{vectoractiondS}
S=\frac{M_P^2}{2}\int d^4 x ~\frac{\beta}{2}\left(\dot{v}_i\dot{v}^i+v_i\nabla^2 v^i+\frac{ca^2(t)}{\beta}v_iv^i\right),
\end{equation}
and the Euler-Lagrange equation of eq.(\ref{vectoractiondS}) is
\begin{equation}\label{vectordS EL eqn}
\left[\beta\left(\frac{d^2}{dt^2}-\nabla^2\right)-ca^2(t)\right]v_i=0.
\end{equation}
By Fourier transform, we find that the solutions are harmonic oscillators with frequency $w_\bold{p}^2=\bold{p}^2-\frac{ca^2(t)}{\beta}$. The canonical momentum conjugate to $v_i$ is then defined by
\begin{equation}
p_{vi}=\frac{\delta S}{\delta\dot{v}_i}=\beta\dot{v}_i
\end{equation}
Since we use the gauge invariant variable to write the action, there is no constraint and the Hamiltonian is
\begin{equation}\label{vectorHdS}
H=\frac{M_P^2}{2}\int d^3x~\frac{p_{vi}p_v^i}{2\beta}-\frac{\beta}{2}v_i\nabla^2v^i-\frac{ca^2(t)}{2}v_iv^i.
\end{equation}

There is a subtle but important difference between de Sitter and Minkowski backgrounds for the scalar modes. In the Minkowski case,  eq.(\ref{vectorH}), the helicity-1 mode is either tachyonic or ghostlike since the sign of $v_iv^i$ is always negative. However, in de Sitter background, one  may choose $c<0$ to render the coefficient of $v_iv^i$ in eq.(\ref{vectorHdS}) to be positive. Nevertheless, as we have argued that generically $c>0$ unless the higher derivative terms dominate, we will not consider this case further. The helicity-1 sector is thus either tachyonic or ghostlike, depending on the sign of $\beta$.

\subsection{Helicity-0 sector}\label{sec:scalarmodedS}
The second order action for helicity-zero modes in dS space is much more complicated. We can use the usual two gauge invariant variables
\begin{align} \label{GI1}
\Phi&=\phi+\dot{B}-\ddot{E}-\frac{1}{t}(B-\dot{E})\nonumber, \\
\Psi&=\psi+\frac{1}{3}\nabla^2E+\frac{1}{t}(B-\dot{E}),
\end{align}
to write the action as
\begin{align}\label{scalaractiondS}
S=&\frac{M_P^2}{2}\int d^4 x~ a^2(t)\left(-6\dot{\Psi}^2-2\Psi\nabla^2\Psi+4\Psi\nabla^2\Phi\right.\nonumber\\
&\left.-12a(t)H\Phi\dot{\Psi}-6a^2(t)H^2\Phi^2\right)\nonumber \\
&+4(\beta+3\alpha)(3\ddot{\Psi}^2+4\dot{\Psi}\nabla^2\dot{\Psi}+2\ddot{\Psi}\nabla^2\Phi) \nonumber \\
&+2(3\beta+8\alpha)(\nabla^2\Psi)^2+2(\beta+2\alpha)(\nabla^2\Phi)^2 \nonumber \\
&-4(\beta+4\alpha)\nabla^2\Psi\nabla^2\Phi+4(\beta+3\alpha)\left[12a^3(t)H^3\Phi\dot{\Psi}\right.\nonumber \\
&+a^2(t)H^2(6\dot{\Psi}^2+6\ddot{\Psi}\Phi+3\dot{\Phi}^2+2\Psi\nabla^2\Psi\nonumber \\
&+7\Phi\nabla^2\Phi-4\Psi\nabla^2\Phi)+2a(t)H(5\dot{\Psi}\nabla^2\Phi+3\ddot{\Psi}\dot{\Phi})\left. \frac{}{}\right].
\end{align}
Again, one can see the action can be reduced to Minkowski case eq.(\ref{scalaraction}) if $a(t)\rightarrow 1$, $H\rightarrow 0$. However, this set of variables is rather unwieldy, so we choose the following pair of gauge invariant variables instead
\begin{align} \label{GI2}
\Phi&=\phi-t(\dot{\psi}+\frac{1}{3}\nabla^2\dot{E}),\nonumber \\
\mathcal{B}&=\nabla^2\left[B-\dot{E}+t\left(\psi+\frac{1}{3}\nabla^2E\right)\right], 
\end{align}
whereupon the action becomes much shorter, i.e.
\begin{align}\label{scalaractiondS2} 
S=&\frac{M_P^2}{2}\int d^4x~ 2(\beta+2\alpha)(\dot{\mathcal{B}}+\nabla^2 \Phi)^2\nonumber \\
&-2a^4H^2 \Phi(\frac{2\mathcal{B}}{aH}+3\Phi)\nonumber \\
&+(\beta+3\alpha)\left\{8aH(\mathcal{B}\nabla^2\Phi+\dot{\mathcal{B}}\dot{\Phi})+16a^3H^3\mathcal{B}\Phi\right.\nonumber \\
&\left.+a^2H^2[8\mathcal{B}^2+8\dot{\mathcal{B}}\Phi+12\dot{\Phi}^2+28(\nabla\Phi)^2]\right\}.
\end{align}
The two set of gauge invariant variables eqs. (\ref{GI1}) and (\ref{GI2}) are related by the following field redefinitions
\begin{equation}
\Phi=\frac{\dot{\Psi}}{a H}+\Phi~,~\mathcal{B}=-\frac{1}{a H}\nabla^2\Psi.
\end{equation}
Using these new set of variables, the canonical momenta can be written as 
\begin{eqnarray}
p_\mathcal{B}&=&(\beta+3\alpha)(8a^2H^2\Phi+8aH\dot{\Phi})\nonumber \\
&& +4(\beta+2\alpha)(\dot{\mathcal{B}}+\nabla^2\Phi),\nonumber \\
p_\Phi&=&(\beta+3\alpha)(24a^2H^2\dot{\Phi}+8aH\dot{\mathcal{B}}),
\end{eqnarray}
and the Hamiltonian thus becomes
\begin{eqnarray} \label{finalbigH}
H&=&\frac{M_P^2}{2}\int d^3x ~\frac{3}{8\beta}\left[p_\mathcal{B}^2-\frac{2p_\mathcal{B}p_\Phi }{3aH}+\frac{(\beta+2\alpha)p_\Phi^2}{6a^2H^2(\beta+3\alpha)}\right]\nonumber \\
&&-\frac{3}{\beta}(p_\mathcal{B}-\frac{p_\Phi}{3aH})[2a^2H^2(\beta+3\alpha)\Phi+(\beta+2\alpha)\nabla^2\Phi]\nonumber \\
&&-8a^3H^3(\beta+3\alpha)\mathcal{B}\left\{\frac{\nabla^2\Phi}{a^2H^2}+\frac{\mathcal{B}}{aH}\right. \nonumber \\
&&\left. +\left[2-\frac{1}{2(\beta+3\alpha)H^2}\right]\Phi\right\}\nonumber \\
&&+\frac{2}{\beta}\Phi\left[2(\beta+2\alpha)(\beta+3\alpha)\nabla^2\nabla^2+12a^4H^4(\beta+3\alpha)^2\right.\nonumber \\
&&\left.+2a^2H^2(\beta+3\alpha)(12\alpha-\beta)\nabla^2+3a^4H^2\beta\right]\Phi.
\end{eqnarray}
To make the dynamics explicit, we perform a final canonical transformation
\begin{eqnarray}
p_\mathcal{B}&\rightarrow& p_\mathcal{B}-\frac{1}{3aH}p_\Phi\nonumber \\
&&+\frac{4}{3aH}[(\beta+2\alpha)\nabla^2+2a^2H^2(\beta+3\alpha)]\mathcal{B}\nonumber \\
\Phi&\rightarrow& \Phi+\frac{1}{3aH}\mathcal{B},
\end{eqnarray}
such that the Hamiltonian becomes
\begin{eqnarray}\label{spin-0H}
H&=&\frac{M_P^2}{2}\int d^3x~ \frac{3}{8\beta}p_\mathcal{B}^2+\frac{p_{\Phi}^2}{48a^2H^2(\beta+3\alpha)}\nonumber \\
&&-\frac{3}{\beta}p_\mathcal{B}\left[(\beta+2\alpha)\nabla^2+2a^2H^2(\beta+3\alpha)\right]\Phi\nonumber \\
&&-\frac{2}{9}\mathcal{B}\left[\frac{(\beta+2\alpha)}{a^2H^2}\nabla^2\nabla^2+2(\beta+3\alpha)\nabla^2\right.\nonumber \\
&&\left.+(3a^2+12a^2H^2(\beta+3\alpha))\right]\mathcal{B}\nonumber \\
&&+\frac{4}{3aH}\mathcal{B}\left[(\beta+2\alpha)\nabla^2\nabla^2+8a^2H^2(\beta+3\alpha)\nabla^2\right.\nonumber \\
&&\left.-12a^4H^4(\beta+3\alpha)\right]\Phi\nonumber \\
&&+\frac{2}{\beta}\Phi\left[2(\beta+2\alpha)(\beta+3\alpha)\nabla^2\nabla^2\right.+12a^4H^4(\beta+3\alpha)^2\nonumber \\
&&\left.+2a^2H^2(\beta+3\alpha)(12\alpha-\beta)\nabla^2+3a^4H^2\beta\right]\Phi.
\end{eqnarray}
Since there is no constraint in this theory, this is the Hamiltonian describing two physical degrees of freedom. Although the instability is not explicitly shown, one can see in some limit the Hamiltonian is unbounded from below. In order to have stable kinetic terms, we require that $(\beta+3\alpha)>0$ and $\beta>0$. On the other hand, in the high frequency limit in Fourier space, one should expect those terms with the highest spatial derivatives to dominate. This requires that $(\beta+2\alpha)<0$ so that the $\mathcal{B}^2$ term  is stable (from the third line in eq.(\ref{spin-0H})), which cannot be satisfied at the same time. We thus conclude the Hamiltonian is unbounded from below with any parameter choice in the high frequency limit.

\section{Stabilization by constraints in de Sitter background}\label{sect::exorcisedS}

\subsection{Helicity-2 sector}
Similar to the Minkowski case, we first rewrite the action (\ref{tensoractiondS}) by introducing a helicity-2 auxiliary tensor field $\lambda_{ij}$
\begin{align}\label{constrainedtensoractiondS}
S=&\frac{M_P^2}{2}\int d^4 x~ \beta\left[(\ddot{E}^\text{TT}_{ij}-\lambda_{ij})^2+2\dot{E}^{\text{TT}ij}\nabla^2\dot{E}^\text{TT}_{ij}\right. ,\nonumber \\
&\left.+(\nabla^2E_{ij}^\text{TT})^2+4\lambda^{ij}\nabla^2E_{ij}\right],\nonumber \\
&+ca^2(t)[(\dot{E}^\text{TT}_{ij})^2+E^{\text{TT}ij}\nabla^2E_{ij}^\text{TT}],
\end{align}
where $\lambda_{ij}$ is transverse traceless and the Lorentz invariance is explicitly broken by $\lambda_{ij}$. Ostrogradski's choice of canonical coordinates is
\begin{align}
E_{ij}\equiv E_{ij}  &\longleftrightarrow \pi^{ij}=2ca^2\dot{E}^{ij}+\beta(-2\dddot{E}^{ij}+2\dot{\lambda}^{ij}+4\nabla^2\dot{E}^{ij}),\nonumber \\
q_{ij}\equiv\dot{E}_{ij} &\longleftrightarrow p^{ij}=2\beta(\ddot{E}^{ij}-\lambda^{ij}),\nonumber \\
\lambda_{ij}\equiv\lambda_{ij}&\longleftrightarrow p^{ij}_\lambda=0,
\end{align}
and the Hamiltonian is
 \begin{align}
H=&\frac{M_P^2}{2}\int d^3 x~ \pi^{ij}q_{ij}+\frac{1}{4\beta}p^{ij}p_{ij}-q^{ij}(ca^2+2\beta\nabla^2)q_{ij}\nonumber \\
&-E^{ij}(\beta\nabla^2\nabla^2+ca^2\nabla^2)E_{ij}+\lambda^{ij}(p_{ij}-4\beta\nabla^2E_{ij}).
\end{align}
The Poisson bracket of a pair of transverse traceless canonical coordinates is identical to their Minkowski counterparts
\begin{equation}
[E_{ij}(\bold{x}),\pi_{kl}(\bold{y})]_{PB}=\hat{\Lambda}_{ij,kl}\delta^{(3)}(\bold{x}-\bold{y}).
\end{equation}
To find the constraints, we apply the Dirac Bracket formalism as usual. It is clear that $p_{\lambda ij}=0$ is a primary constraint, and the rest of the (transverse and traceless) constraints of this theory are generated by the consistency relation
\begin{align}
\varphi_1&:p_{\lambda ij}=0,\nonumber \\
\varphi_2&:p_{ij}-4\beta\nabla^2E_{ij}\approx0,\nonumber \\
\varphi_3&:\pi_{ij}-2ca^2q_{ij}\approx0,\nonumber \\
\varphi_4&:2(\beta\nabla^2\nabla^2+ca^2\nabla^2)E_{ij}-\frac{ca^2}{\beta}p_{ij}\nonumber \\
&+2(-ca^2+2\beta\nabla^2)\lambda_{ij}\approx 0.
\end{align}
Armed with these, we can use $\varphi_1$, $\varphi_4$ to eliminate the degree of freedom $(\lambda, p_\lambda)$, and use $\varphi_2$, $\varphi_3$ to eliminate $(q, p)$. The coefficients in the action (\ref{constrainedtensoractiondS}) are again chosen such that there are at least four constraints in the theory and there is no $\nabla^2$ in $\varphi_3$ which will generate nonlocal terms in the reduced Hamiltonian. 

Using the constraints, $(q_{ij}, p_{ij})$ can be written as follow
\begin{align}
q_{ij}&=\frac{\pi_{ij}}{2ca^2},\nonumber \\
p_{ij}&=4\beta\nabla^2E_{ij},
\end{align}
and the reduced Hamiltonian becomes
 \begin{align}\label{reducedHdS}
H_R=&\frac{M_P^2}{2}\int d^3~ x\frac{1}{4c^2a^4}\pi^{ij}(ca^2-2\beta\nabla^2)\pi_{ij}\nonumber \\
&+E^{ij}(-ca^2\nabla^2+3\beta\nabla^2\nabla^2)E_{ij},
\end{align}
which is positive definite if $\beta>0$, $c>0$.

\subsection{Helicity-1 sector}
In section \ref{sec:vectormodedS}, we showed that the helicity-1 modes are only stable if $\beta > 0$, $c<0$.  However, our imposition of the constraints to restore stability of the helicity-2 sector requires that $c>0$ in addition to the usual arguments on subdominant higher derivative terms. We thus choose $c>0$ and remove the unstable helicity-1 modes altogether as follows.
Similar to the Minkowski case, we modify the action (\ref{vectoractiondS}) by introducing a helicity-1 field $\lambda_i$
\begin{equation}\label{mdvectoractionds}
S=\frac{M_P^2}{2}\int d^4 x ~\frac{\beta}{2}\left[(\dot{v}_i-\lambda_i)^2+v_i\nabla^2 v^i+\frac{ca^2}{\beta}v_iv^i\right].
\end{equation}
Ostrogradski's choice of canonical coordinates is
\begin{align}
v_i\equiv v_i&\longleftrightarrow p_{v}^i=\beta(\dot{v}^i-\lambda^i),\nonumber\\
\lambda_i\equiv \lambda_i&\longleftrightarrow p_{\lambda}^i=0,
\end{align}
and the Hamiltonian is 
\begin{align}
H=\frac{M_P^2}{2}\int d^3 x ~\frac{p_v^ip_{vi}}{2\beta}+p_v^i\lambda_i-\frac{\beta}{2}v_i\nabla^2v^i-\frac{ca^2}{2}v_iv^i.
\end{align}
There are four constraints in the theory, which can be found as
\begin{align}
\varphi_1&: p^i_\lambda=0\nonumber \\
\varphi_2&: p^i_v\approx0\nonumber \\
\varphi_3&: ca^2v^i+\beta\nabla^2v^i\approx 0\nonumber \\
\varphi_4&:\frac{ca^2p^i_v}{\beta}+\nabla^2p^i_v+ca^2\lambda^i+\beta\nabla^2\lambda^i\approx 0.
\end{align}
If we use the four constraints to eliminate $(v_i, p^i_v)$, $(\lambda_i, p_\lambda^i)$, the physical phase space will be zero dimensional and the reduced Hamiltonian vanishes.

\subsection{Helicity-0 sector}
Finally, we deal with the helicity-0 instability. We modify the action (\ref{scalaractiondS2}) by introducing a helicity-0 field $\lambda$
\begin{align}\label{mdscalaractiondS}
S=&\frac{M_P^2}{2}\int d^4x ~2(\beta+2\alpha)(\dot{\mathcal{B}}-\lambda+\nabla^2 \Phi)^2\nonumber \\
&-2a^4H^2 \Phi\left(\frac{2\mathcal{B}}{aH}+3\Phi\right)\nonumber \\
&+(\beta+3\alpha)\left[8aH(\mathcal{B}\nabla^2\Phi+\dot{\mathcal{B}}\dot{\Phi}-\lambda \dot{\Phi})\nonumber \right.\\
&+16a^3H^3\mathcal{B}\Phi+28a^2H^2\Phi\nabla^2\Phi\nonumber \\
&+a^2H^2\left.(8\mathcal{B}^2+8\dot{\mathcal{B}}\Phi-8\lambda \Phi+12\dot{\Phi}^2)\right].
\end{align}
As now must be familiar, the canonical coordinates are
\begin{align}
p_\lambda=&0\nonumber \\
p_\Phi=&(\beta+3\alpha)(24a^2H^2\dot{\Phi}+8aH\dot{\mathcal{B}}-8aH\lambda)\nonumber\\
p_\mathcal{B}=&4(\beta+2\alpha)(\dot{\mathcal{B}}-\lambda+\nabla^2\Phi)\nonumber \\
&+(\beta+3\alpha)(8a^2H^2\Phi+8aH\dot{\Phi})
\end{align}
and the Hamiltonian is 
\begin{align}
H=&\frac{M_P^2}{2}\int d^3x~ \frac{3}{8\beta}\left[p_\mathcal{B}^2-\frac{2p_\mathcal{B}p_\Phi }{3aH}+\frac{(\beta+2\alpha)p_\Phi^2}{6a^2H^2(\beta+3\alpha)}\right]\nonumber \\
&-\frac{3}{\beta}(p_\mathcal{B}-\frac{p_\Phi}{3aH})[2a^2H^2(\beta+3\alpha)\Phi+(\beta+2\alpha)\nabla^2\Phi]\nonumber \\
&+\frac{2}{\beta}\Phi\left[2(\beta+2\alpha)(\beta+3\alpha)\nabla^2\nabla^2+12a^4H^4(\beta+3\alpha)^2\right.\nonumber \\
&\left.+2a^2H^2(\beta+3\alpha)(12\alpha-\beta)\nabla^2+3a^4H^2\beta\right]\Phi\nonumber \\
&-8a^3H^3(\beta+3\alpha)\mathcal{B}\left\{\frac{\nabla^2\Phi}{a^2H^2}+\left[2-\frac{1}{2(\beta+3\alpha)H^2}\right]\Phi\right.\nonumber \\
&\left. +\frac{\mathcal{B}}{aH}\right\}+\lambda p_\mathcal{B}.
\end{align}
There are four constraints in the theory, which can be found as
\begin{align}
\varphi_1&: p_\lambda=0\nonumber \\
\varphi_2&: p_\mathcal{B}\approx0\nonumber \\
\varphi_3&: \frac{2\mathcal{B}}{aH}+\frac{\nabla^2\Phi}{a^2H^2}+\left[2-\frac{1}{2H^2(\beta+3\alpha)}\right]\Phi\approx 0\nonumber \\
\varphi_4&:F(\lambda, \cdots)\approx 0.
\end{align}
Applying these constraints to remove $\lambda$ and $\mathcal{B}$ pairs, we obtain the reduced Hamiltonian
\begin{align}\label{HRdSspin0}
H_R=&\frac{M_P^2}{2}\int d^3x~ \frac{(\beta+2\alpha)}{16a^2H^2\beta(\beta+3\alpha)}p_\Phi^2+2(\beta+3\alpha)(\nabla \Phi)^2\nonumber \\
&-2a^2[1+6(\beta+3\alpha)H^2](\partial\Phi)^2+2a^4H^2\Phi^2\nonumber \\
&+\left[\frac{8a^4H^4(\beta+3\alpha)(2\beta+5\alpha)}{(\beta+2\alpha)}+\frac{a^4}{2(\beta+3\alpha)}\right]\Phi^2.
\end{align}
If we require $H_R>0$, which means every term in eq.(\ref{HRdSspin0}) needs to be positive definite, there are two possibilities:

\begin{itemize}
\item{$\alpha\leq 0$, $\beta+3\alpha>0$ which guarantees $\beta>0$.}
\item{$\alpha>0$, $\beta+2\alpha>0$, if we also require that the helicity-1, 2 modes are stable, the second condition becomes $\beta>0$.}
\end{itemize}
Choosing either possibility will result in a stable helicity-0 sector.

\section{Conclusion}\label{sect:conclusion}
We investigate the instabilities in higher derivative gravity models with quadratic curvature invariant $R^2$, $R^{\mu\nu}R_{\mu\nu}$ by expanding action to the quadratic level of metric fluctuation around Minkowski/de Sitter background. We show how the instabilities in the helicity-0, 1, 2 sectors can be removed by some choices of additional constraints. With help of the constraints, the degrees of freedom are reduced from \emph{two} helicity-0, \emph{two} helicity-1, and \emph{four} helicity-2 to \emph{one} helicity-0, \emph{zero} helicity-1, and \emph{two} helicity-2 modes. The fact that the phase space has to be reduced -- i.e. it is impossible to modify the theory via constraints such that the instabilities are ``made stable'' -- is an expression of the theorem proven in \cite{Chen:2012au} that Ostrogradski's instability can only be removed if the original theory's phase space is reduced.

We emphasize that adding constraints to remove instabilities is only valid in the linear theory. A full non-linear extension of this methodology is beyond the scope of this paper, and we have made no attempt at a covariant formalism. However, even in the linear theory, some features of a stable higher derivative gravity can be gleaned. First, it is clear that a general higher derivative theory which is stable and possesses the desirable renormalization properties breaks Lorentz invariance.  Indeed, the ``stabilized'' theory has the form of a low energy effective limit of a  Lorentz violating, much like that of Ho\u{r}ava gravity.

Second, the stable higher derivative theory has no helicity-1 modes, at least in the Minkowski case since this mode is unstable in the original theory and hence need to be removed. The de Sitter case is less clear-cut -- the helicity-1 sector may be made stable by the curvature term although we have chosen to remove it to be consistent with the stability of the helicity-0 sector. It will be interesting to check whether this result can be extended to the full non-linear regime. We will leave this for future work.

\acknowledgments

We would like to thank Alex Vikman, Ignacy Sawicki, Andrew Tolley, Daniel Baumann, Matteo Fasielo, Kirk Hinterbichler, Pau Figueras, and Helvi Witek for useful conversations. 

\appendix

\section{Quantization of higher derivative theory} \label{sect:appendix1}
In this section, we use a higher derivative scalar field theory to demonstrate the subtleties of a quantum higher derivative theory. We begin with 
\begin{equation}\label{A1}
S=\int d^4x~ \frac{1}{2}\phi\Box \phi+\frac{\sigma}{2M^2}(\Box \phi)^2-\frac{m^2\phi^2}{2},
\end{equation}
 where $\Box$ is d'Alembert operator, $M$, $m$ are constants with mass dimension $1$ and $\sigma=\pm1$. The Euler-Lagrange equation is 
\begin{equation}
\Box \phi+\frac{\sigma}{M^2}\Box \Box \phi-m^2\phi=0,
\end{equation} 
by Fourier transform, the solution is a set of harmonic oscillators with frequency 
\begin{equation}
w_p^2-p^2=\frac{-M^2\pm M^2\sqrt{1+\frac{4\sigma m^2}{M^2}}}{2\sigma}.
\end{equation}
We can see that there are two frequencies correspond to each $p$, which means the theory has two degrees of freedom. To simplify the calculation, we can take $m=0$ and one of the d.o.f. thus becomes massless. 
We can also take $\sigma=-1$, which makes the other d.o.f ghostlike ($\sigma=1$ would instead make it a tachyonic ghost). 
We can thus denote the frequencies by
\begin{align}
w_p^2&=p^2\nonumber \\
v_p^2&=p^2+M^2.
\end{align}  
To describe the theory in the Hamiltonian picture we need to first define the canonical variables
\begin{align}
q_1=\phi &\Leftrightarrow p_1=\frac{\delta S}{\delta \dot{\phi}}\nonumber \\
q_2=\dot{\phi} &\Leftrightarrow p_2=-\frac{\ddot{\phi}}{M^2}.
\end{align}
Since there is no constraint in the theory, the Hamiltonian is 
\begin{eqnarray} \label{h}
H&=&\int d^3x~ p_1q_2-\frac{M^2p_2^2}{2}+q_1(-\frac{1}{2}\nabla^2+\frac{1}{2M^2}\nabla^2\nabla^2)q_1\nonumber \\
&&+q_2(-\frac{1}{2}+\frac{1}{M^2}\nabla^2)q_2.
\end{eqnarray}
To quantize the theory, we write $q_1$, $q_2$, $p_1$, $p_2$ as linear combinations of the two pairs of creation and annihilation operators $(a_p^\dagger, a_p)$, $(b_p^\dagger, b_p)$
\begin{align}\label{equation}
q_1=\int \frac{d^3p}{(2\pi)^3}~&\left[ \frac{1}{\sqrt{2w_p}}(a_pe^{ip\cdot x}+a_p^\dagger e^{-ip\cdot x})\right.\nonumber \\
+&\left. \frac{i}{\sqrt{2v_p}}(b_p^\dagger e^{ip\cdot x}-b_pe^{-ip\cdot x})\right]\nonumber \\
q_2=\int \frac{d^3p}{(2\pi)^3}~&\left[ (-i)\sqrt{\frac{w_p}{2}}(a_pe^{ip\cdot x}-a_p^\dagger e^{-ip\cdot x})\right.\nonumber \\
+&\left. \sqrt{\frac{v_p}{2}}(b_p^\dagger e^{ip\cdot x}+b_pe^{-ip\cdot x})\right]\nonumber \\
p_1=\int \frac{d^3p}{(2\pi)^3}~&\left[\frac{ (-i)}{M^2}\sqrt{\frac{w_pv_p^4}{2}}(a_pe^{ip\cdot x}-a_p^\dagger e^{-ip\cdot x})\right.\nonumber \\
+&\left. \frac{1}{M^2}\sqrt{\frac{v_pw_p^4}{2}}(b_p^\dagger e^{ip\cdot x}+b_pe^{-ip\cdot x})\right]\nonumber \\
p_2=\int \frac{d^3p}{(2\pi)^3}~&\left[ \frac{1}{M^2}\sqrt{\frac{w^3_p}{2}}(a_pe^{ip\cdot x}+a_p^\dagger e^{-ip\cdot x})\right.\nonumber \\
+&\left. \frac{i}{M^2}\sqrt{\frac{v^3}{2}}(b_p^\dagger e^{ip\cdot x}-b_pe^{-ip\cdot x})\right].
\end{align}
The coefficients of creation and annihilation operators are chosen in the way that the commutators
\begin{align}
[q_1(x), p_1(y)]=[q_2(x), p_2(y)]=i\delta^{(3)}(x-y)
\end{align}
are consistent with the usual commutator relation
\begin{align}
[a_p, a_k^\dagger]=[b_p, b_k^\dagger]=(2\pi)^3\delta^{(3)}(p-k),
\end{align}
with all other possible commutators vanishing. The other thing one should notice is each canonical variable is combination of two degrees of freedom, the two d.o.f. vibrate at different frequencies, i.e. in the Heisenberg picture, $a_p\rightarrow a_pe^{-iw_pt}$, $b_p\rightarrow b_pe^{iv_pt}$. With all these information, we can substitute eq.(\ref{equation}) into the Hamiltonian (\ref{h}), after some work we find the Hamiltonian to be 
\begin{align}
H=\int\frac{d^3p}{(2\pi)^3}~&w_p\left[a_p^\dagger a_p+\frac{1}{2}(2\pi)^3\delta^{(3)}(0)\right]\nonumber \\
-&v_p\left[b_p^\dagger b_p+\frac{1}{2}(2\pi)^3\delta^{(3)}(0)\right],
\end{align}
where $w_p=\sqrt{p^2}$ and $v_p=\sqrt{p^2+M^2}$. One can see while $a_p^\dagger$ creates a massless particle with positive energy, $b_p^\dagger$ create a massive particle with negative energy, thus the theory has a massive ghost. One can always redefine $b_p\equiv b_p^\dagger$, and the new $b_p^\dagger$ will create a massive particle with positive energy but saddled with a negative norm.

\section{Equivalence of Ostrogradski's formalism and auxiliary field method}
In this appendix, we will use eq.(\ref{A1}) with $\sigma=-1$ and $m=0$, as a toy-model to show the equivalence between Ostrogradski's formalism of higher derivative theory and the auxiliary field method used in the literature (e.g. \cite{Stelle:1976gc}). In the auxiliary field method, the action with one higher derivative scalar field
\begin{equation}
S=\int d^4x~ \frac{1}{2}\phi\Box \phi-\frac{1}{2M^2}(\Box \phi)^2,
\end{equation}
is equivalent to the action with two standard scalar fields
\begin{equation}
S=\int d^4x~ \frac{1}{2}\phi\Box \phi-\frac{1}{2M^2}(\Box \phi)^2+\frac{1}{2M^2}\left[\Box\phi+\frac{M^2(\lambda-\phi)}{2}\right]^2.
\end{equation}
The action can be reduced to
\begin{equation}
S=\int d^4x~ \frac{1}{2}\lambda\Box \phi+\frac{M^2}{8}(\lambda-\phi)^2,
\end{equation}
and diagonalized as
\begin{equation}\label{B2}
S=\int d^4x~ \frac{1}{2}\Phi\Box \Phi-\frac{1}{2}\Psi\Box \Psi+\frac{M^2}{2}\Psi^2,
\end{equation}
where $\phi=\Phi-\Psi$ and $\lambda=\Phi+\Psi$. The action (\ref{B2}) describes a healthy massless scalar field with a massive ghostlike scalar field. The conjugate momenta and the Hamiltonian of the system can be easily written as 
\begin{align}\label{B3}
p_\Phi&=\dot{\Phi}\nonumber \\
p_{\Psi}&=-\dot{\Psi} \nonumber \\
H&=\int d^3x~ \frac{p_\Phi^2}{2}+\frac{(\partial \Phi)^2}{2}-\left[\frac{p_\Psi^2}{2}+\frac{(\partial \Psi)^2}{2}+\frac{M^2}{2}\Psi^2\right].
\end{align}
On the other hand, Ostrogradski's formalism leads to the Hamiltonian (\ref{h}), which is linearly dependent on $p_1$
\begin{align} 
H=&\int d^3x~ p_1q_2+q_1\left(-\frac{1}{2}\nabla^2+\frac{1}{2M^2}\nabla^2\nabla^2\right)q_1\nonumber \\
&-\frac{M^2p_2^2}{2}+q_2\left(-\frac{1}{2}+\frac{1}{M^2}\nabla^2\right)q_2,
\end{align}
which can be diagonalized by the following canonical transformation
\begin{align}
q_1=&\Phi+\Psi\nonumber \\
q_2=&p_\Phi-p_\Psi\nonumber \\
p_1=&p_\Phi-\frac{\nabla^2}{M^2}(p_\Phi-p_\Psi)\nonumber \\
p_2=&\Psi-\frac{\nabla^2}{M^2}(\Phi+\Psi).
\end{align}
The final Hamiltonian becomes
\begin{equation}
H=\int d^3x ~\frac{p_\Phi^2}{2}+\frac{(\partial \Phi)^2}{2}-\left[\frac{p_\Psi^2}{2}+\frac{(\partial \Psi)^2}{2}+\frac{M^2}{2}\Psi^2\right],
\end{equation}
which is same as eq.(\ref{B3}). Hence we have shown that Ostrogradski's formalism is equivalent to the auxiliary field method up to some canonical transformation.

\bibliography{mybib}
\end{document}